\def\PR{{Phys.~Rev.~}}
\def\PRL{{ Phys.~Rev.~Lett.~}}

\def\sVEC{\text{\small{VEC~}}}

\documentclass[aps,prl,superscriptaddress,a4paper,twocolumn]{revtex4-1}
\usepackage{amsmath}
\usepackage{color}
\usepackage{floatrow}
\usepackage{amssymb}
\usepackage{graphicx}

\begin{document}
\title{Ta-Nb-Mo-W refractory high-entropy alloys:  anomalous ordering behavior \\ and its intriguing electronic origin}
\author{Prashant Singh}
\email{prashant@ameslab.gov}
\affiliation{Ames Laboratory, U.S. Department of Energy, Iowa State University, Ames, Iowa 50011 USA}
\author{A. V. Smirnov}
\affiliation{Ames Laboratory, U.S. Department of Energy, Iowa State University, Ames, Iowa 50011 USA}
\author{Duane D. Johnson}
\email{ddj@iastate.edu, ddj@ameslab.gov}
\affiliation{Ames Laboratory, U.S. Department of Energy, Iowa State University, Ames, Iowa 50011 USA}
\affiliation{Department of Materials Science \& Engineering, Iowa State University, Ames, Iowa 50011 USA}

\begin{abstract} 
From electronic-structure-based thermodynamic linear-response, we establish chemical ordering behavior in complex solid solutions versus how Gibbs' space is traversed -- applying it on prototype refractory A2 Ta-Nb-Mo-W high-entropy alloys. 
Near ideal stoichiometry, this alloy has anomalous, intricate chemical ordering tendencies, with long-ranged chemical interactions that produce competing short-range order (SRO) with a crossover to spinodal segregation.
This atypical SRO arises from canonical band behavior that, with alloying, create features near the Fermi-surface (well-defined even with disorder) that change to simple commensurate SRO with (un)filling of these states. Our results reveal how complexity and competing electronic effects control ordering in these alloys.
\end{abstract}
\maketitle


Demand and pursuit for materials with  high thermal stability and good high-temperature mechanical response have never faded for practical applications and scientific interest \cite{Ritchie2011,Buban2006,Wu2010,Chen2015}. Multi-component High-entropy alloys (HEA), proposed to stabilize simple phases using maximum entropy \cite{Yeh1,Zhang2014,Gludovatz2014,Gludovatz2016}, has shown good merit and has led to improved mechanical behavior \cite{Cantor2004,Yao2014,Deng2015,Ma2015,Li2016,AdMater2017,Singh2018,Buenconsejo2009}.  However, it is valid only for a fraction of alloy composition space \cite{Guo2011}.  Moreover, non-equiatomic compositions greatly increases the design space for tailoring phase stability and associated mechanical behavior in more general complex solid-solution alloys (``HEA''), without sacrificing much entropy  \cite{Li2016,AdMater2017,Singh2018}. Empirical rules extended from binaries have been used to guide the HEA stability, experimentally focusing on size-effect and thermodynamics. Hume-Rothery's 15\% size-effect rule was shown to arise from an alloy's electronic structure  (a difference in bandwidths of the alloying elements) \cite{Pinski1991}. With  competing elemental sizes, it has been supposed that this effect in HEA is large, e.g., in Cantor alloys \cite{Cantor2004, Gludovatz2016}, which has proven to be incorrect  \cite{ZhangSRO}.  More recently,  tailoring composition from ``metastability-engineering"  \cite{Herrera2011,Steinmetz2013,Grassel2000,Sun2013,Marteleur2012} and entropy  \cite{Yeh1} have been successfully joined \cite{Li2016,AdMater2017,Singh2018}. 

For near-equiatomic HEAs with N-components (\text{N}$\,\ge\,$4) \cite{Yeh1,Cantor2004}, the design strategy has been to stabilize the random solid-solution in simple crystal lattices \cite{Tsai2017,Feuerbacher3,Takeuchi4} (retarding formation of intermetallics  \cite{Takeuchi4,Senkov2011_1,GMS2015}) and attempt to find specific electronic, thermodynamic, and microstructural properties \cite{Yeh1, Chuang7,Wang8, Hemphill9, Kozelj10}  for multifarious  applications \cite{Gludovatz2014,Tsai2008,Kozelj10}.  Experiments indicate that bcc HEA, in particular refractory elements with  their high-melting points, could exhibit  stable microstructure at high temperatures (T) with large heat-softening resistance, even better than conventional Ni-based superalloys \cite{Senkov2011_1,Senkov2010,Zou2015,Senkov2011,Senkov2013}. 
What is missing, however, is a first-principle guide that combines electronic, thermodynamic, and mechanical alloying effects, as we do here for Ta-Nb-Mo-W refractory alloys, where competing electronic effects govern the behavior across the entire 4-D composition space, which cannot be captured by empirical rules or methods. 

{\par} To identify candidates, CALPHAD methods, or similar approaches, using thermodynamic databases mostly have been applied \cite{Senkov2015,Miracle2017}.  Predicting properties of complex solid-solutions remains challenging from electronic-structure methods. Results are available for relative stability estimates and competing long-range order (LRO) states \cite{Huhn2013,Kormann2017}, but provide limited understanding and are restricted in composition. For example, relative global stability (formation enthalpies, $\Delta E_f$) does not address local chemical stability (short-range order, SRO), which affects experiment but is difficult to assess due to high-T annealing required in refractory systems and sluggish diffusion \cite{Zhang2014_1,Tsai2013}. Hence, a robust electronic theory of alloying is critical to identify thermodynamic and electronic origins for properties; here, we reveal Fermi-surface features that dictate key properties that are relevant to all refractory systems \cite{Singh2018}.

{\par}To that end, we calculate $\Delta E_f$ and SRO for any arbitrary HEAs and identify the mechanism at play. We use thermodynamic linear-response \cite{Singh2015} to predict  $\alpha_{\mu\nu}$({\bf{k}};T) pair-correlations with $\frac{1}{2}\text{N}(\text{N}-1)$ $\mu$-$\nu$ atom pairs. 
$\Delta E_f$ and SRO are numerically evaluated using all-electron KKR (Green's function) and the coherent-potential approximation (CPA) to handle chemical disorder \cite{JohnsonCPA}, with screened-CPA is used to address charge correlations from Friedel screening \cite{JP1993}.  Details of the calculations are in the Supplemental Material.  We exemplified this quantitative agreement with experiment for Al$_x$CrFeCoNi, including the range of two-phase coexistence \cite{Singh2015,AS2016}. 

{\par} Here we applied the theory to A2 TaNbMoW (TNMW) and detail the competing electronic origins for ordering in this complex refractory alloy, which sensitive to direction traversed in $\{c_{\mu}\}$ space. TNMW displays anomalous chemical ordering sensitive to T and $\{c_{\mu}\}$, with real-space SRO parameters that are long-ranged only at equiatomic case. 
From ${\bf k}$-space linear-response theory \cite{SJP1994,AJP1995,AJPS1996}, both short- and long-ranged SRO are accurately represented and its origin is linked directly with band-filling (valence electron count, VEC); Fermi surfaces \cite{AJPS1996,GS1983}, hybridization \cite{Pinski1991}, and van Hove states \cite{Clark1995}.

\begin{figure}[t!]
{\caption {(Color online) Gibbs' tetrahedron (top) sighted along Nb-Mo-Ta face to rear W vertex. Plane (white-dash) through TaNbMoW is parallel to Nb-W-Ta plane. Binaries {\footnotesize {(A-NbMo; B-TaW; C-TaMo; D-NbW; E-TaNb; F-MoW)}} identify paths to TaNbMoW (VEC=5.5): isoelectronic (ISO1 A-B [red]; ISO2 C-D [blue]), and band-filling (BF) E-F [green]. E$^{A2}_f(0\,K)$ vs. $x$ for paths (bottom), with binary mixtures favorable, except $x\sim0.5$. Measured TaW E$^{A2}_f$ is marked (see also Fig. S1).}  \label{fig1}}
{\includegraphics[scale=0.17]{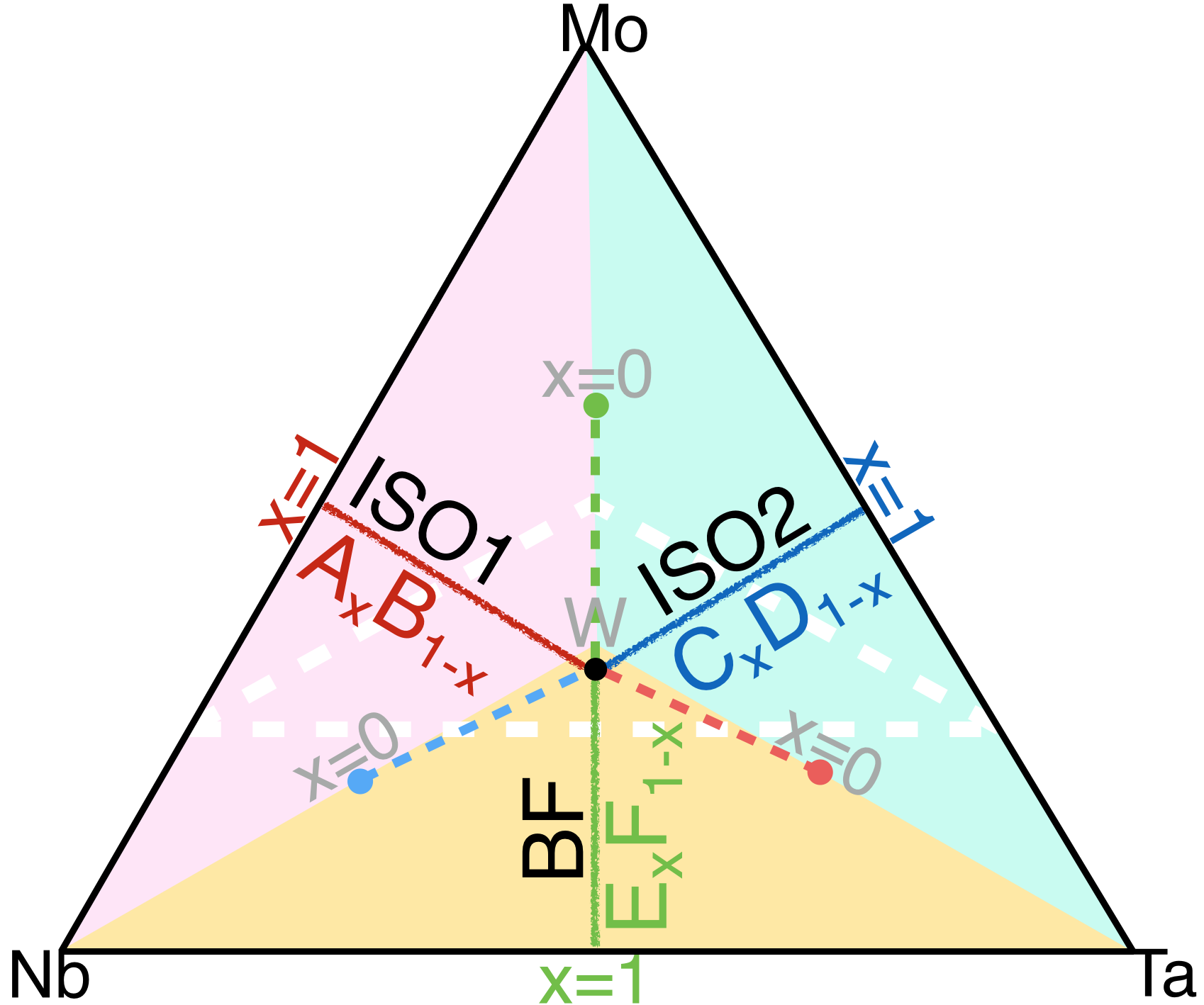} \\ \includegraphics[scale=0.3]{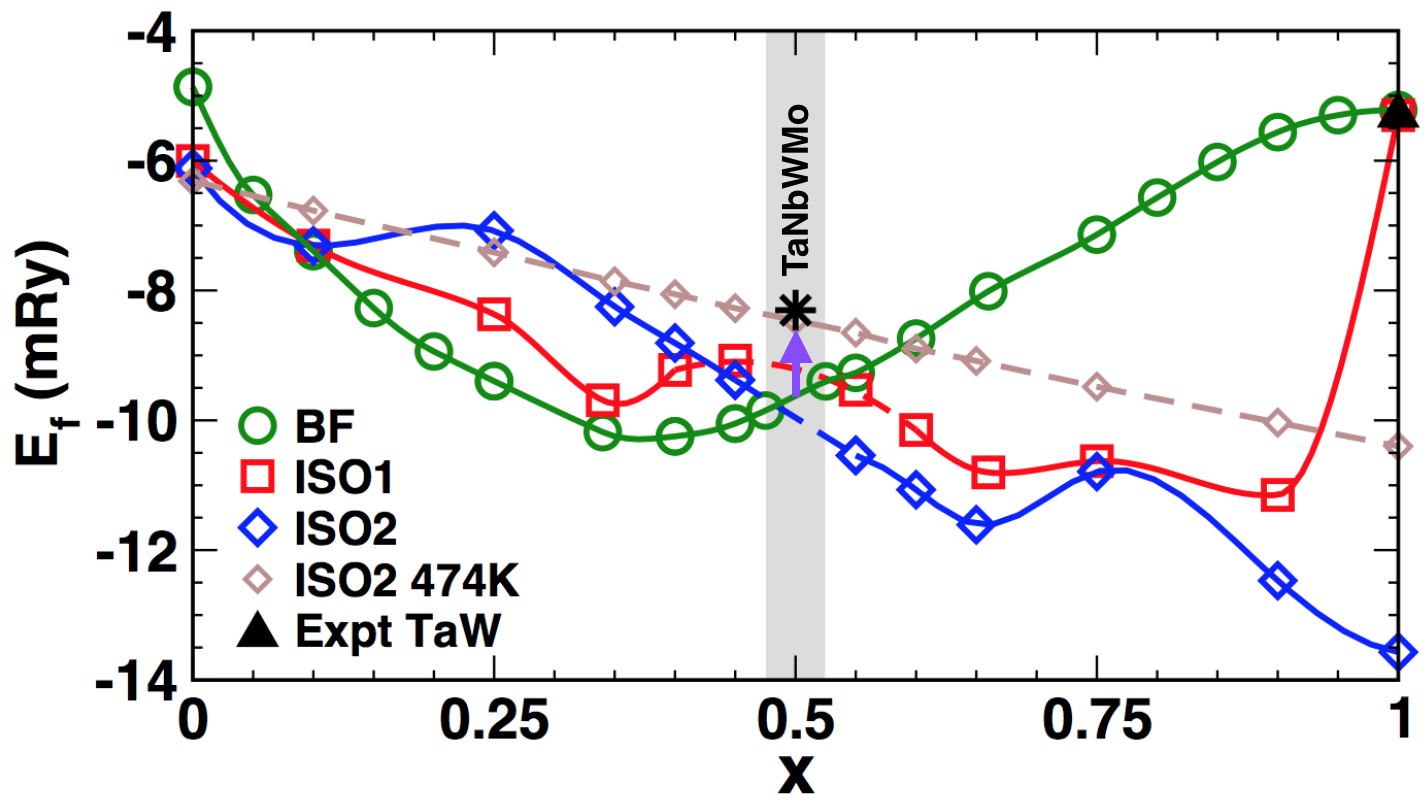} }
\end{figure}

{\par}To set the stage, Ta,Nb (Mo,W) are in group 5 (6), specifying their VEC.  In  Gibbs' space (N$=$4 vertices, Fig.~\ref{fig1}), there are 6 possible $\mu$-$\nu$ pairs (edges) requiring 2 binaries to make a HEA, giving three main paths for TNMW, which pass through $\{c_{\nu}\}$\,=$\,\frac{1}{4}$ for atom-type $\nu$.  This equiatomic alloy is a unique point due to canonical band behavior of group 5 (6) elements, where, for example, Cr bands, scaled by bandwidth, become that of Mo and W. We will focus first on the SRO versus VEC that is dependent on alloying and dispersion.

{\par} The fluctuation-dissipation theorem connects SRO pair correlations to responses from induced compositional variations $\{\delta c_{\nu}^{i}\}$ at site $i$ from small inhomogeneous chemical potentials $\{\delta \nu_{\mu}^{j}\}$ \cite{SJP1994,AJP1995,AJPS1996}: \begin{small}$q^{ij}_{\mu\nu} = \delta c_{\nu}^{i}/\delta \nu_{\mu}^{j}= \left< x^{i}_{\mu}x^{j}_{\nu}\right> - \left<x^{i}_{\mu}\right>\left<x^{j}_{\nu}\right> \equiv \alpha^{ij}_{\mu\nu} \, c_{\mu}(\delta_{\mu\nu}-c_{\nu})$\end{small}.
Here, occupation variables $x^{i}_{\mu}\,$=$\,1\,(0)$ if site $i$ is (is not) atom-type $\mu$, and $\{x^{i}_{\mu}\}$ represents any configuration; by thermal averaging, $c^{i}_{\mu}$=$\left<x^{i}_{\mu}\right>$. 
Warren-Cowley SRO parameters $\alpha^{ij}_{\mu\nu}$ are normalized so that pair probabilities are 
$P^{ij}_{\mu\nu} = c^{i}_{\mu} c^{j}_{\nu} [1-\alpha^{ij}_{\mu\nu}]$.
We have short-range ordering for $\alpha<$0 [clustering for $\alpha>$0] with bounds
 $-{[\min(c_{\mu},c_{\nu})]^2}{(c_{\mu}c_{\nu})^{-1}} \leq \alpha_{\mu\nu}^{i\neq j} \leq 1$.
With the homogeneous state ($c_{\alpha}^{i}\,$=$\,c_{\alpha}$\,$\forall~i$) as reference, $\alpha_{\mu\nu}({\bf{k}};T)$  allows us to assess all ordering modes simultaneously, as the Fourier transform uses the symmetry of the underlying Bravais lattice, just as done for phonons.   
In terms of concentration (Fourier) waves, we calculate the non-singular portion of inverse correlation matrix [$\{\mu,\nu\} \in 1,\text{N}$-$1$] relative to the N$^{th}$  ``host'', 
{\par} \begin{small} 
\begin{equation} \label{SROeq}
\left[q^{-1}({\bf k};T)\right]_{\mu\nu} = \left( \frac{\delta_{\mu\nu}}{c_{\mu}}+\frac{1}{c_{N}}\right) -  \beta S_{\mu\nu}^{(2)}({\bf k};T) ,
\end{equation}
	\end{small}
\noindent  where $\beta\equiv(\text{k}_{\text{B}} \text{T})^{-1}$ and $\text{k}_{\text{B}}$ is Boltzmann's constant, and 
$q^{-1}_{\mu\nu}({\bf k};T)  \, c_{\mu}(\delta_{\mu\nu}-c_{\nu}) \equiv \alpha^{-1}_{\mu\nu}({\bf k};T)$.
$S_{\mu\nu}^{(2)}({\bf k})$ is the second-variation with respect to $c_{\mu}^{i}$,$c_{\nu}^{j}$, or curvature, of the KKR-CPA grand potential; physically, it is a chemical stability matrix. 
Equation (\ref{SROeq}) is exact \cite{GS1983,AJP1995,AJPS1996,Evans1979}.
For any approximate $\bar{S}_{\mu\nu}^{(2)}({\bf k})$, like single-site CPA, we require that 
    $S_{\mu\nu}^{(2)}({\bf k}) = \bar{S}_{\mu\nu}^{(2)}({\bf k})-\Lambda_{\mu\nu}$
to enforce intensity conservation, where Onsager cavity fields satisfy (via Newton-Raphson)
 \begin{small} \begin{equation} \label{LAMeq}
\Lambda(T)_{\mu\nu} = \sum_{\gamma}\frac{1}{\Omega_{BZ}} \int_{BZ} d{\bf k} \: \bar{S}_{\mu\gamma}^{(2)}({\bf k};T) \alpha_{\gamma\nu}({\bf k};T) . 
\end{equation} \end{small}
For  T$\rightarrow$$\infty$ or more exact $\bar{S}_{\mu\nu}^{(2)}$, $\Lambda_{\mu\nu}\rightarrow0$, otherwise it improves the temperature scale and corrects the topology of mean-field phase diagrams \cite{Teck2011}, even for a singe-site.

{\par} Thermodynamically $S_{\mu\nu}^{(2)}({\bf k})$  are pair-interchange energies, with contributions from all orders in sites (not just pairs) via the reference; they reveal the unstable (Fourier) modes with wavevector {\bf k}$_{o}$, and can identify origins for phase transitions. Small (large) positive $S_{\mu\nu}^{(2)}({\bf k})$ values give the high-energy, short-lived (low-energy, long-lived) fluctuations observable in diffuse scattering. For ${\bf k}_{o}\,$=$\,\Gamma$ (long-wavelength) mode, the alloy is unstable to decomposition.  Where $\alpha^{-1}_{\mu\nu}({\bf k}_{o};T_{sp})=0$ is the instability to this mode at  spinodal $T_{sp}$, an estimate for critical temperatures \cite{Singh2015}. 
We evaluate S$_{\mu\nu}^{(2)}$ in a \emph{band-energy-only} approximation, as double-counting terms are small by the force theorem in the homogeneous state \cite{Singh2015,AJPS1996}.

\begin{figure*}[!]
\floatbox[{\capbeside\thisfloatsetup{capbesideposition={right},capbesidewidth=4.4cm}}]{figure}[\FBwidth]
{\caption{(Color online) For TaNbMoW at an optimized lattice $a_{o}\,$=$\,6.113\,a.u.$, S$^{(2)}_{\mu\nu}({\bf k};T)$~(a)~and $\alpha_{\mu\nu}({\bf k};T)$~(b) along high-symmetry lines. Ta-Mo is dominant, with $T_{sp}\,$=$\,1080\,$K from ${\bf k}_{o}\,$=$\,\Gamma$, indicative of decomposition. S$^{(2)}_{\text{TaMo}}({\bf k}_o;1.85T_{sp})$ gives instability at ${\bf k}_o$$\approx$$\frac{1}{2}$(P-H), an incommensurate SRO, as typically for FS ($`2k_{F}$') nesting. Crossover occurs near $1.57T_{sp}\approx$1700\,K.
}\label{fig2} }
{\includegraphics[scale=0.35]{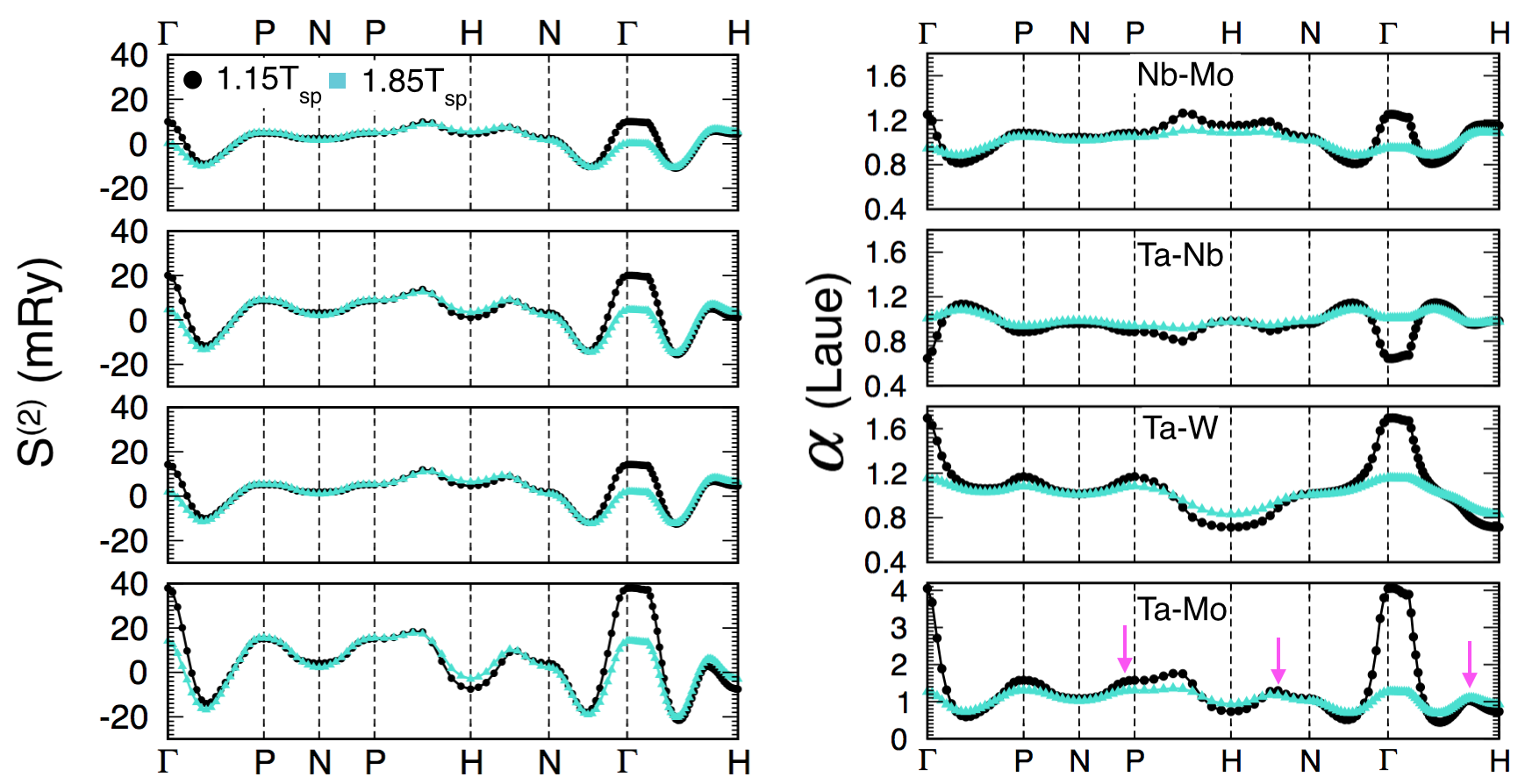}}
\end{figure*}

{\par}S$_{\mu\nu}^{(2)}$ is a susceptibility \cite{GS1983,AJPS1996} having energy $\epsilon$- and species $\mu$-dependent matrix elements, two Fermi factors $f(\epsilon,T)$ for (un)filling states, and a convolution of the KKR-CPA scattering path operator, $\tau_{LL'}({\bf k};\epsilon)$ 
that embodies all electronic-structure effects in a HEA, including dispersion, defined by the Bloch spectral function (BSF), i.e., $A({\bf k};\epsilon)\,$=$\,-\pi^{-1}\mathcal{I}\it{m}\,\tau({\bf k};\epsilon)$. For ordered dispersion, $A({\bf k};\epsilon)\,$=$\,\delta(\epsilon -  \epsilon_{{\bf k},s})$; the trace of the loci of BSF peaks at the Fermi energy $\epsilon_F$ defines the Fermi surface (FS).  With disorder, $\delta$-functions broaden and shift in {\bf k} for a given $\epsilon$, decreasing lifetimes in a state and increasing resistivity; the loci of BSF peaks at  $\epsilon_F$ still defines the FS: ${\bf k}$ is a good (not exact) quantum number  if widths are well-defined on the scale of the Brillouin zone.  If FS dictate SRO, then a convolution of $\tau({\bf k};\epsilon_F) \tau({\bf k}+{\bf q};\epsilon_F)$ is relevant \cite{GS1983,Moss1964}, where nested sheets (displaced by a single ${\bf q}_{\text{s}}$) produce a convolution (enhanced by disorder broadening) giving constructive diffuse intensity at points along high-symmetry lines (geometrically given by overlaps of spheres with radii `${2k_F}$' =$|{\bf q}_{\text{s}}|$).

{\par}In Fig.~\ref{fig2} for TNMW, we plot S$_{\mu\nu}^{(2)}$({\bf k};T) and $\alpha_{\mu\nu}({\bf k})$ for the largest four pairs.  Above $T_{sp}$, Ta-Mo is the dominant pair for interchange energies and SRO. At high temperatures (as in annealing or quenching experiments), clearly S$^{(2)}_{\text{TaMo}}({\bf k}_o;1.85T_{sp})$ drives an instability at ${\bf k}_o$$\approx$$\frac{1}{2}$(P-H), an incommensurate (long-period) SRO; a peak just larger than that at $\Gamma$. Other pairs have similar behavior.  Upon cooling, the instability in $\alpha_{\mu\nu}({\bf k}_o;1.15T_{sp})$, driven by S$_{\text{TaMo}}^{(2)}({\bf k}_o;1.15T_{sp})$, has crossed over to ${\bf k}_o$=$\Gamma$, which occurred at $1.57T_{sp}\approx1700\,$K. The absolute instability is reached at $T_{sp}(\Gamma)\,$=$\,1080\,$K, a third of melting.

\begin{figure}[b!]
\includegraphics[scale=0.3]{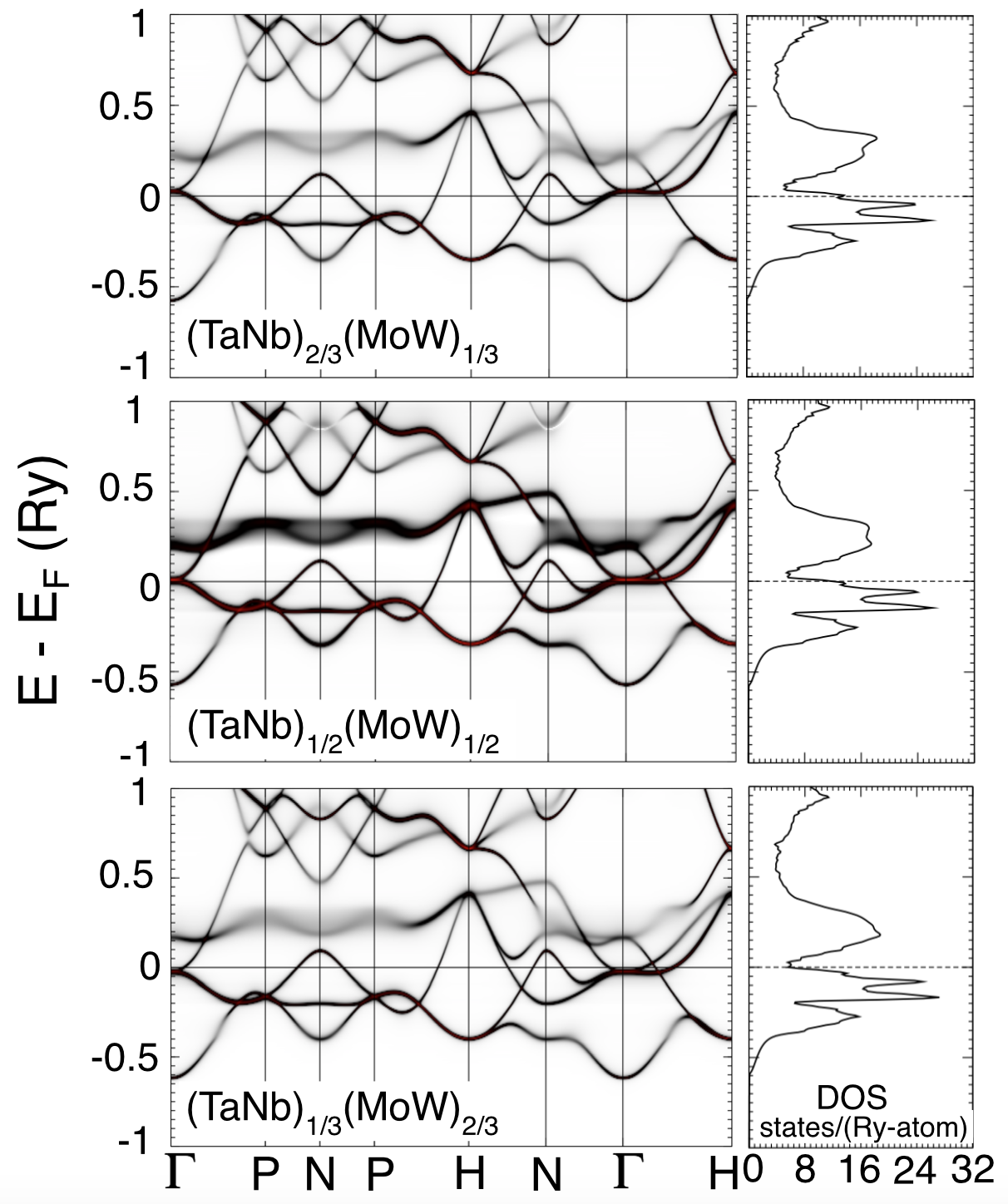} 
\caption {(Color online) Dispersion (BSF) and DOS (relative to $\epsilon_{\text{F}}$=0) for (TaNb)$_{x}$(MoW)$_{1-x}$. VEC varies by $-\frac{1}{6}, 0, +\frac{1}{6}$ with $x$=$\frac{2}{3},\frac{1}{2},\frac{1}{3}$. Flat states near $\Gamma$ are $35$, $0$, and $-25$\,mRy from $\epsilon_{\text{F}}$ vs. $x$. All binary BSF are compared to HEAs in Fig.~S1.}
\label{fig3}
\end{figure}

\begin{figure*}[!]
\floatbox[{\capbeside\thisfloatsetup{capbesideposition={right},capbesidewidth=5.2cm}}]{figure}[\FBwidth]
{\caption {(Color online)  For \begin{small}(TaNb)$_{x}$(WMo)$_{1-x}$\end{small} with $a_o$($x$=$\frac{1}{2}$), S$^{(2)}_{\text{TaMo}}({\bf k};T)$~(a), $\alpha_{\text{TaMo}}({\bf k};T)$~(b), and S$^{n}_{\text{TaMo}}$~(c),(d)  for $n^{th}$-neighbor shell. [All 6 pairs are compared in Fig.~S5.] Near $x$=$\frac{1}{2}$ Fermi-surface nesting yields long-ranged S$^{n}$ and crossover of SRO due to a flat dispersion near $\Gamma$, Fig.~\ref{fig3}. Otherwise S$^{n}_{\text{TaMo}}$ are short-ranged ($n$\,$<$\,10). SRO thus depends strongly on band filling near $x$=$\frac{1}{2}$, as it is B32-type ({\bf k}$_{o}$=P$\{ \frac{1}{2}\frac{1}{2}\frac{1}{2} \}$) at $x$=$\frac{2}{3}$ and B2-type ({\bf k}$_{o}$=H$\{111\}$) at $x$=$\frac{1}{3}$, and no temperature dependence (for clarity only 1.15$T_{sp}$ results are shown). $T_{sp}$ and {\bf k}$_{o}$ is provided in upper left of each panel in (b).} \label{fig4} }
{\includegraphics[scale=0.27]{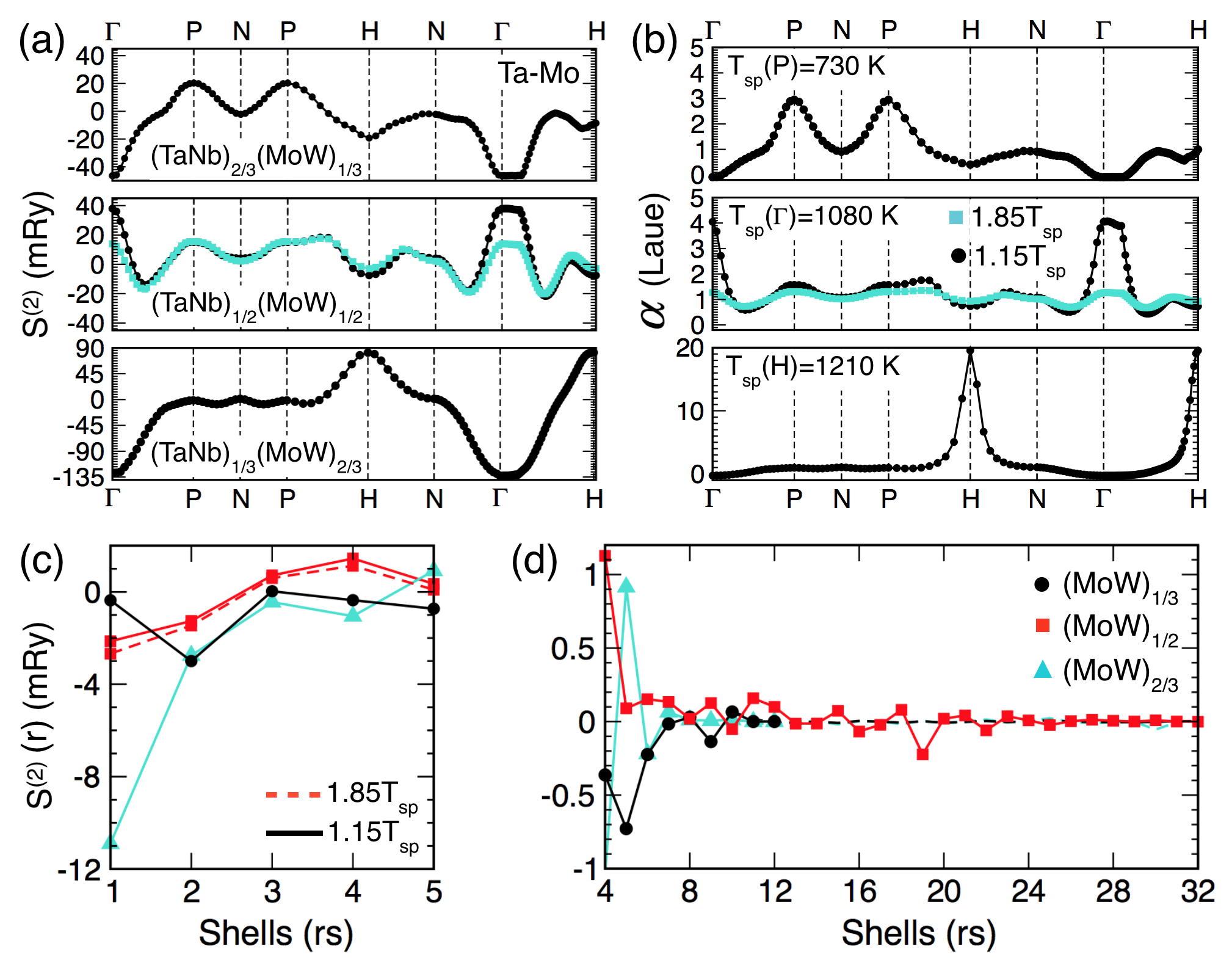}}
\end{figure*}

{\par}In Fig.~\ref{fig2}, the temperature dependence of S$_{\mu\nu}^{(2)}$({\bf k};T) and subsidiary peaks in $\alpha_{\mu\nu}({\bf k})$, e.g., for $\frac{1}{2}$P-H (strong), $\frac{1}{2}$H-N (medium) and $\frac{3}{4}$$\Gamma$-H (weak), indicate presence of band-filling and FS-nesting effects.  To verify, we study (TaNb)$_{x}$(WMo)$_{1-x}$ along the band-filling path (Fig.~\ref{fig1})  and follow changes in the dispersion (Fig.~\ref{fig3}), density of states (DOS, Fig.~\ref{fig3}), and SRO (Fig.~\ref{fig4}). From Fig.~\ref{fig3}, the DOS of ideal HEA($x\,$=$\,\frac{1}{2}$) shows that $\epsilon_F$ sits just below the $d$-band pseudogap with some (non)bonding states left unfilled.  With $x\,$=$\,\frac{2}{3}$($\,\frac{1}{3}$), \sVEC varies by $-\frac{1}{6}$($+\frac{1}{6}$), and disorder broadening above $\epsilon_F$ noticeably changes, along with $s$-bandwidth. Otherwise, a single feature at $\epsilon_F$ stands out:  a flat (low-dispersion) $d$-band along $\Gamma$-H.   Increasing (decreasing) \sVEC (de)populates  this dispersion and affects stability and SRO. The HEA($x\,$=$\,\frac{1}{3}$) should be (and is) more stable, with a stronger commensurate SRO and higher $T_{sp}$.

{\par} In Fig.~\ref{fig4}, SRO of the alloys are compared for the dominant Ta-Mo pair.   
Unfilling the flat dispersion ($x\,$=$\,\frac{2}{3}$) leads to S$_{\mu\nu}^{(2)}$({\bf k};T) with instability at P $\{ \frac{1}{2}\frac{1}{2}\frac{1}{2} \}$, Fig.~\ref{fig4}(a), and a  weak commensurate B32-type SRO, Fig.~\ref{fig4}(b), with $T_{sp}$ of $730\,$K.  Filling these states ($x\,$=$\,\frac{1}{3}$) gives instability at H $\{111\}$, and a stronger B2-type SRO with $T_{sp}$ of $1210\,$K. 
Second, a T-dependent S$_{\mu\nu}^{(2)}$({\bf k};T) occurs only at $x\,$=$\,\frac{1}{2}$. This flat dispersion at $\epsilon_F$  imparts a sensitivity to (un)filling of states from the Fermi terms in the susceptibility, i.e., \begin{small}$\frac{f({\epsilon};T)-f({\epsilon}';T)}{({\epsilon}-{\epsilon}')}$\end{small}; with states (un)filled, there can be no overt temperature dependence.

\begin{figure}[b]
\centering
\includegraphics[scale=0.3]{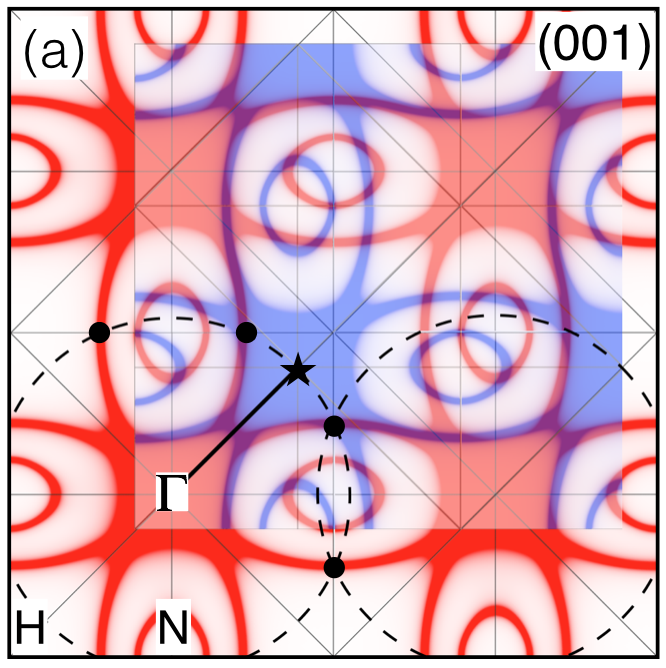}
\includegraphics[scale=0.3]{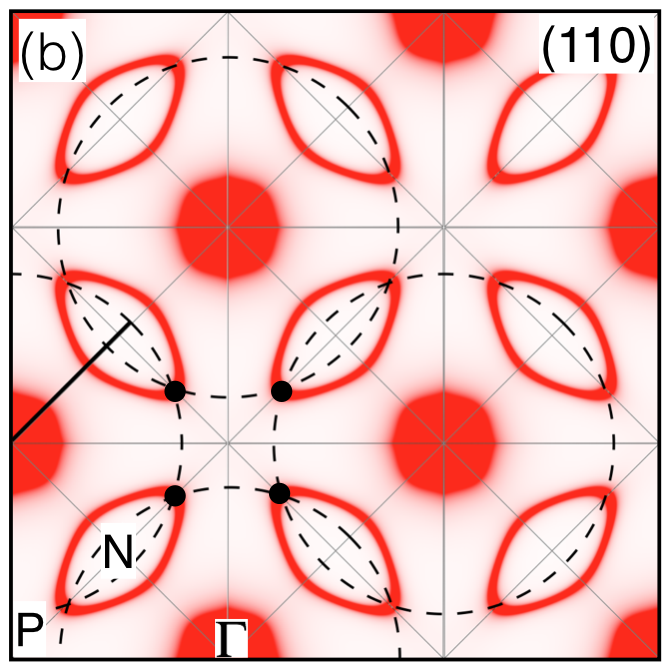}
\caption {(Color online)  TaNbMoW Fermi surfaces (red) in (a)~(001) and (b)~(110) planes. Relevant convolution of states is given as shifted ($\Gamma$$-$$\star$) blue FS in (a), with ${\bf q}_{s}$=$\frac{3}{4}\,$($\Gamma$-\text{H}).  A set of intensity peaks [filled circles in (a),(b)]  expected from nesting are shown at overlaps of `2k$_F$' spheres (dashed circles), matching incommensurate intensities in Fig.~\ref{fig2}.}
\label{fig5}
\end{figure}

{\par}Notably, these flat states are the same as in A2 Cr that exhibit ${\bf q}_s\,$$\sim$$\,0.97\,\Gamma$-H between flat, square $\Gamma$-centered hole states and H-centered electron states and give well-known incommensurate antiferromagnetism \cite{Callaway1981}. Adding solute,  antiferromagnetism becomes commensurate with coexisting superconductivity \cite{Johnson.MRS.1988}, just like `Fe-As' superconductors!  Interestingly, the canonical bands of Cr can be scaled by bandwidth to produce those of Mo or W \cite{Skriver1981}; as such these Cr Fermi-surface pockets decrease around $\Gamma$ and enlarge around H, and, with alloying, they look like those in Fig.~\ref{fig5}. So, the canonical bands with alloying and hybridization produce all the observed effects, making the ideal HEA a special point in Gibbs' space.

{\par} It remains to explain the subsidiary features in the SRO in the ideal HEA  (Fig.~\ref{fig2}), and its  unique behavior versus temperature.
From above, the FS is playing a key role, and, if nesting is involved, it makes the real-space S$_{\mu\nu}^{n}$ long-ranged. S$_{\text{TaMo}}^{n}$ is plotted versus neighbor shells in Fig.~\ref{fig4}(c,d) for each case. For VEC$\,>5.5$ (or $\,<5.5$), S$_{\text{TaMo}}^{n}$ is short-ranged, and dominated by shells 1-6, with no T dependence. For the ideal HEA,  S$_{\text{TaMo}}^{n}$ is very long-ranged ($n>24$ shells), see Fig.~\ref{fig4} and Table S1.
The Fermi surface is well-defined  in the A2 Brillouin zone (Fig.~\ref{fig5}). For ${\bf q}_{s}\,$=$\,\frac{3}{4}\,$($\Gamma$-\text{H}) 
there is a FS convolution (a remnant from elemental dispersion), with equal contributions for shifts along any $\Gamma$-\text{H}. Any FS-driven peaks from convolutions is anticipated by drawing circles of radius $|{\bf q}_{s}|$ from all $\Gamma$ points \cite{GS1983}, and intensity occur on these arcs, e.g., at $\frac{3}{4}\,$($\Gamma$-\text{H}).  Larger intensities occur where more circles overlap, e.g., at $\frac{1}{2}$(N-H) in Fig.~\ref{fig2}(a), and $\frac{1}{2}$(P-H). These intensities manifest as subsidiary peaks (arrows in Fig.~\ref{fig2}(b)), e.g., for $\alpha_{\text{TaMo}}$({\bf k}). Such effects can  dominate, as in Cu-Pd long-period order \cite{GS1983}, Cu-Ni-Zn Heusler order \cite{AJP1995,AJPS1996}, or CuPt van Hove-driven L1$_1$ order \cite{Clark1995}.

{\par} For isoelectronic cases in Fig.~\ref{fig1}, one might anticipate the TNMW incommensurate SRO along the path. For A2 TaW, for example, our calculated E$^{A2}_f$ agrees very well with experiment (Fig.~\ref{fig1} and S1), and LRO states [high to low: B32, B2, and L2$_1$] agree with band methods. While B2 is much lower than A2, L2$_1$ is slightly lower than B2, as indicated by the SRO (Fig.~S4).  Binaries have  dispersion like TNMW but with less broadening (Fig.~S2). Well away from $\{c_{\nu}\}$\,=$\,\frac{1}{4}$, where maximum complexity occurs,  the endpoint binaries all have calculated B2 SRO and LRO. For ISO1 case with $x\,$=$\,1$$\rightarrow$$\frac{2}{3}$$\rightarrow$$\frac{1}{2}$ (or $x\,$=$\,0$$\rightarrow$$\frac{1}{3}$$\rightarrow$$\frac{1}{2}$), SRO transitions from $\text{H}$, to $\Gamma$, to $\Gamma$ with competing $\text{H}$ and $\text{P}$ (T-dep. $S^{(2)}$), Fig.~S5. For this isoelectronic line, complexity increases and peaks at TNMW, making Fermi-surface effects operative.

{\par} Notably, the S$_{\mu\nu}^{(2)}({\bf k}_o;T)$ eigenvectors ${\bf e}({\bf k}_o)$ indicate the expected LRO unit cell and site probabilities (Fig.~S6)  after symmetry breaking \cite{Singh2015}. At 1.85$T_{sp}$, ${\bf e}({\bf k}_o)$ at $\frac{1}{2}$(P-H)  gives a mode that is near degenerate with commensurate B2 and B32. Hence, with B2+B32 high-T SRO, a sample should exhibit this competing ordering if quenched from above $1.57T_{sp}$;  if annealed, it should  spinodal decompose dominated by Ta-Mo pairs (Fig.~S4). However, decomposition requires good diffusion, whereas B2+B32 SRO is already established and lower in free energy than a segregating alloy. A B2+B32 mixed state below $T_{sp}$ is then possible -- an  explanation for why no clear order has been observed in this HEA \cite{Senkov2011}. 
Thus, a crossover from standard to anomalous ordering arises from canonical bands, band-filling and FS effects.
Recent generalized perturbation method (GPM) $r$-space results \cite{Kormann2017} predict commensurate SRO with a B2+B32 mixed groundstate, with FS origin dismissed as impossible. We both find similar results above the phase boundary, but, like our SRO, GPM interactions are not generally applicable below a phases boundary. The KKR-CPA and high-T linear-response was applied across the entire compositions space and quantitatively revealed the origin for changing properties and ordering behavior.

{\par}  In conclusion, combining formation enthalpy (both above and below a phase boundary) with thermodynamic linear-response applied to complex multiple-principal element alloys reveals directly the chemical ordering modes (short-range order and expected long-range order) and its electronic origins  -- an ideal approach for predictive design.
We established that refractory A2 TaNbMoW has a complex ordering near ideal stoichiometry that change rapidly to simple commensurate ordering with (un)filling of states or decreased disorder broadening, depending on how Gibbs' space is traversed. This behavior results from canonical bands (and band-filling) and key Fermi-surface features established by alloying (both relevant in general refractory systems \cite{Singh2018}), effects not found in simulations based on semi-empirical potentials or real-space, but often used recently for materials design.

\vspace{1em}
\acknowledgments
 Work supported by the U.S. Department of Energy (DOE), Office of Science, Basic Energy Sciences, Materials Science \& Engineering Division. Research was performed at ISU and Ames Laboratory, which is operated by ISU for the U.S. DOE under contract DE-AC02-07CH11358.

\end{document}